\documentclass[a4paper,graphicx,12pt]{article}
\usepackage{graphics}
\usepackage{cite}
\usepackage{authblk}
\usepackage{epsfig}
\usepackage{graphicx}

\begin{titlepage}
\title{\bf Surface Magnetism of  Fe(001), Co(001) and Ni(001)}
\author[1]{Priyadarshini Parida $^1$, Biplab Ganguli\footnote{Corresponding Author\\Email:biplabg@nitrkl.ac.in (Biplab Ganguli), abhijit.mookerjee61@gmail.com (Abhijit Mookerjee)}} 
\author[2,3,4]{Abhijit Mookerjee}
\affil[1]{Department of Physics, National Institute of Technology Rourkela - 769008, India.}
\affil[2]{Department of Physics and Materials Science, S.N. Bose National Centre for Basic Sciences, JD-III Salt Lake, Kolkata - 700098, India.}
\affil[3]{Distinguished Visiting Professor, Presidency University, College Street, Kolkata, India.}
\affil[4]{Visiting Professor, Lady Brabourne College, Suhrawardy Avenue, Kolkata, India.}
\date{}
\end{titlepage}

\begin{document}
\maketitle
\begin{abstract}
The Augmented Space Formalism coupled with Recursion method and Density Functional Theory based Tight-Binding Linear Muffin-Tin Orbitals have been applied for a first principles calculation of surface electronic and magnetic properties of body centered cubic Fe(001) and face centered cubic Co(001) and  Ni(001). Nine atomic layers have been  studied to see the trend of change in these properties from surface into the bulk. Surface magnetic moment has been found to be higher than that of the bulk and  in different layers below, magnetic moments show Friedel oscillations in agreement with other studies. Work functions of these systems have been found to agree with experimental values. We propose this real space technique to be  suitable for the study of localized physical properties like surface layers and it is also suitable for the study rough surfaces and interfaces.   
\end{abstract}


\section{Introduction}
Itinerant Fermi electrons are responsible for the magnetic ordering in transition metals. The main contribution to magnetism in transition metals comes from the d-orbital valence electrons. These properties are strongly influenced the by local environment like surfaces and interfaces. Surface magnetic properties are quite different from those in the  bulk. At the surface the co-ordination number and symmetry decrease. Magnetic moment get enhanced at the surface compared to its bulk value \cite{huda, monodeep, ohnishi, freeman, wu, freeman1,li,  wimmer, eriksson, eriksson1, alden, niklasson, biplab,  wang}. Since magnetic properties depend on detailed electronic structure, atomic arrangement and the composition, therefore it is necessary to investigate detailed electronic structure of a surface to understand the surface effects on magnetic properties. It is known \cite{wu, eriksson, niklasson, freeman2} that enhancement of surface magnetism arises because of band narrowing at the surface due to decrease in the coordination number and symmetry. As we go down into the bulk from the topmost layer, density of states and magnetic moment approach their bulk values. The contribution of s- and p-orbital is negligible in these materials.

Number of computational methods like FPLAPW \cite{ohnishi,freeman,wu,freeman1,li,wimmer}, LMTO \cite{eriksson,eriksson1}, TB-LMTO Green's function method \cite{alden,niklasson}, TB-LMTO supercell calculation \cite{huda, biplab}, TB-LMTO recursion method \cite{monodeep} and LCAO method \cite{wang, wang1} have been adopted for the study of surface magnetism. All these methods except TB-LMTO recursion method are $\vec{k}$-space methods. Therefore they are applicable only in cases where periodicity is maintained. A real space method can also be applied to cases where periodicity breaks down. At a surface, periodicity breaks down in the perpendicular direction. All the $\vec{k}$-space methods are still applicable to a surface as long as we embed the surface in a super-cell, with a layer of vacuum and it maintains periodicity in 2-dimensions. But a surface and an interface are often rough. Roughness of a surface is created due to vacancy defects at surface layers while a layer is formed during an epitaxial growth on a substrate. Similarly due to diffusion of surface atoms into the substrate and vice versa, a rough interface is formed. Therefore, a surface and an interface can be treated as disordered systems.

For more accurate calculations of physical properties at a surface it is also necessary to take into account charge leakage into the vacuum. This is in general carried out by considering one or more layers of empty sphere above the surface. This consideration also breaks the periodicity in perpendicular direction. In such cases a real space techniques are more appropriate in dealing with such disorder at the surface.

Augmented Space Formalism (ASF) \cite{mookerjee, mookerjee1}, introduced originally for the study of binary alloys is extended to deal with a rough surfaces \cite{biplab}.  Therefore we have coupled ASF  with the recursion method \cite{haydock}, (ASR) \cite{tanusri, rudra} to study surface magnetism. Therefore, ASF with recursion method is a good tool to study the localized properties of disordered systems considering all the configurations. But recursion works on a localized basis so it can be combined with TB-LMTO \cite{andersen} formalism. We have applied this method here and considered only smooth surface. In our subsequent works we shall include cases of roughnes at various layers of a surface and interfaces between magnetic surface and matal subtrate.    

We have taken two layers of empty spheres on top of the nine layers of Fe, Co as well as Ni to take care of charge leakage. Because of reduced co-ordination number at the surface, surface atoms are weakly bound as compared to those in the bulk and so the interatomic separation is expected to be slightly different from  its bulk. Therefore lattice relaxation is necessary. In a previous TB-LMTO-ASR calculation for Fe (001) surface \cite{biplab}, surface dilatation was carried out. In this communication we consider lattice relaxation of surface layers.

Since work function is an important quantity characterizing a surface which is also often measured experimentally, therefore its calculation is important for comparison with experimental value. The experimental measurement of surface magnetic moment is not available for most of the conductors for comparison with calculated value. Our calculated  work functions agree quite well with  experimental values \cite{michael}.

\section{Computational Methods}

ASF \cite{mookerjee, mookerjee1} deals with configuration averaging of the random variables on the same footing as quantum mechanical averaging by augmenting the Hilbert space spanned by the wavefunctions with a disorder or configuration space spanned by the different realizations of the random Hamiltonian. In the recursion method successive recursions of any initial vector $|u_0\rangle$ on the Hamiltonian gives  new bases \{ $|u_n\rangle$\} which are mutually orthogonal to each other.
The recurrence relation is given by,

\begin{eqnarray*}
H |u_n\rangle = a_n|u_n\rangle + b_{n+1}|u_{n+1}\rangle +b_n |u_{n-1}\rangle 
\end{eqnarray*}

where, $a_n$ and $b_n$ are the recursion coefficients which are the matrix elements of the
tri-diagonal matrix in the new representation. In this basis, the resolvent $(zI-H)^{-1}$ can be written in the form of a continued fraction. The continued fraction is terminated after a finite number of steps using a suitable terminator. We have used the Luchini and Nex terminator \cite{luchini} in our work.

 But recursion works on a localized basis. Therefore a localized basis like TB-LMTO is quite suitable. Hence it is combined with TB-LMTO formalism. The full ASF hamiltonian in TB-LMTO basis \cite{tanusri} is given by,
\begin{eqnarray*}
H_{RL,R'L'}^\alpha &=& \hat{C}_{RL}\delta_{RR'}\delta_{LL'} + \hat{\triangle}_{RL} S_{RL,R'L'}^\alpha \hat{\triangle}_{R'L'}\\
\hat{C}_{RL}&=& C_{RL}^A n_R + C_{RL}^B (1-n_R)\\
\hat{\triangle}_{RL}&=& \triangle_{RL}^A n_R + \triangle_{RL}^B (1-n_R)
\end{eqnarray*}
Here $R$ is the lattice sites and $L=(lm)$ are the orbitals indices. For transition metal $l<2$. $C_{RL}^A$, $C_{RL}^B$, $\triangle_{RL}^A$ and $\triangle_{RL}^B$ are the TB-LMTO potential parameters of the constituents $A$ and $B$ of the alloy. $n_R$ are the local site occupation variables which randomly takes value 1 and 0 according to whether the site is occupied by an $A$ atom or not. Combining these three method, the effective hamiltonian is derived, which is used to carry out the surface properties. From the effective hamiltonian, local density of states and local magnetic moment are calculated. When both types of atoms in a binary alloy are taken of identical types, a substitutional disordered system approaches to an ordered system. Therefore ASF can be applied to a smooth surface if we take surface atom in place of empty sphere also in the ASF formalism. We have adopted this procedure in the present case for our smooth surface.

We have chosen nine layers of bcc Fe(001), fcc Co(001) and fcc Ni(001). On top of these, we have introduced two layers of empty spheres which contain charge but no atoms. This empty spheres will take care of the charge leakage into the vacuum.  The potential parameters are generated from TB-LMTO within local spin density approximation (LSDA) using Barth and Hedin exchange correlation potential. Wave equations are solved by the scalar-relativistic calculations. The lattice relaxation for top most layer in three system is carried out by minimizing the total energy. It is found that for bcc Fe(001) the lattice relaxation is  5\%. The relaxations obtained for fcc Co(001) and fcc Ni(001) are respectively 16\% and 9\%. We have used seven shell augmented space calculation and nine steps of recursion. 

\section{Results and Discussions}

\begin{figure}[h!]
\centering
\includegraphics[width=8cm,height=12cm]{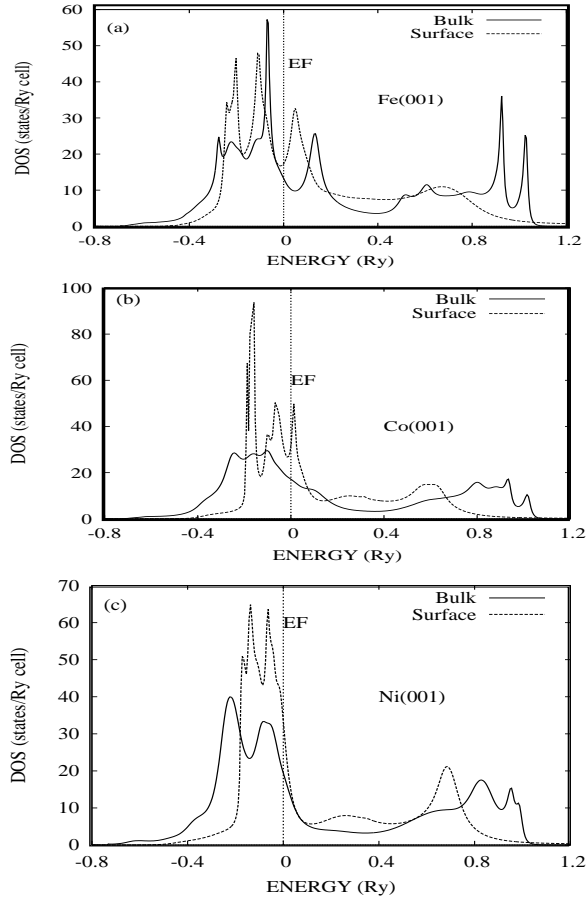}
\caption{Comparison of total density of states for bulk and surface (top most layer). Vertical line at zero shows fermi level.}
\end{figure}
As discussed earlier in the introduction that energy bands of the surface states near Fermi level get narrower due to weakening of interaction by symmetry breaking and reduction in co-ordination number. This is indeed observed in the figure 1 of the total density of states for all the three systems. Similar picture arise in the spin resolved DOS as shown in the figure 2. Apart from narrowing of band there is change in number of spin-up surface states compared to the bulk states  at the Fermi level. The amount of change is maximum in the case of Fe(001) and negligible in the case of Ni(001). But in all the three cases, there is significant change in the spin-down states. The splitting of spin-up and spin-down states near Fermi level is maximum for Fe(001) and least for  Ni(001). This is expected because magnetic moment of Fe is maximum and it is least for Ni. The magnetic moment is directly related to the amount of splitting in spin up and spin down states.
 
\begin{figure}[h]
\centering
\includegraphics[width=8cm,height=12cm]{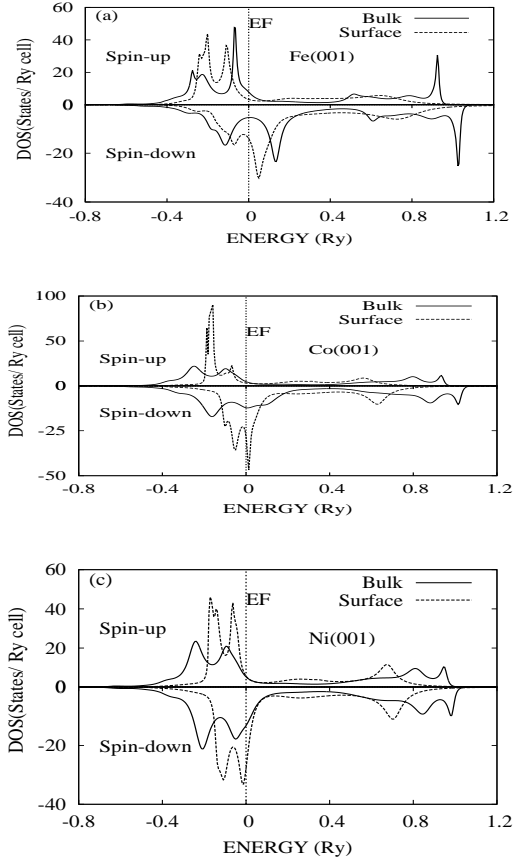}
\caption{Comparison of spin resolved density of states for bulk and surface (top most layer). Vertical line at zero shows fermi level.}
\end{figure}

It is expected that Fermi energy of the surface electrons is lower compared to the bulk. This is because of leakage of surface charge to the vacuum. Due to charge leakage, the electron density decreases. Hence the total surface electronic energy as well as the repulsion  between surface atoms is also reduced compared to the bulk. This causes decrement in fermi energy at the surface. This energy is found to be -0.13 Ry for Fe(001). But after five layers, it is -0.07 Ry which is closed to the bulk value(-0.08 Ry). Similarly the energy at the surface of Co(001) is -0.27 Ry and after four layers it is -0.06 Ry, which is again closed to the bulk value (-0.08 Ry). For Ni(001), they are respectively  -0.24 Ry and -0.08 Ry  after four layers, which is closed to bulk value(-0.09 Ry). The bulk fermi energies are calculated directly from TB-LMTO method \cite{andersen}.

\begin{figure}[h]
\centering
\includegraphics[width=14cm,height=8cm]{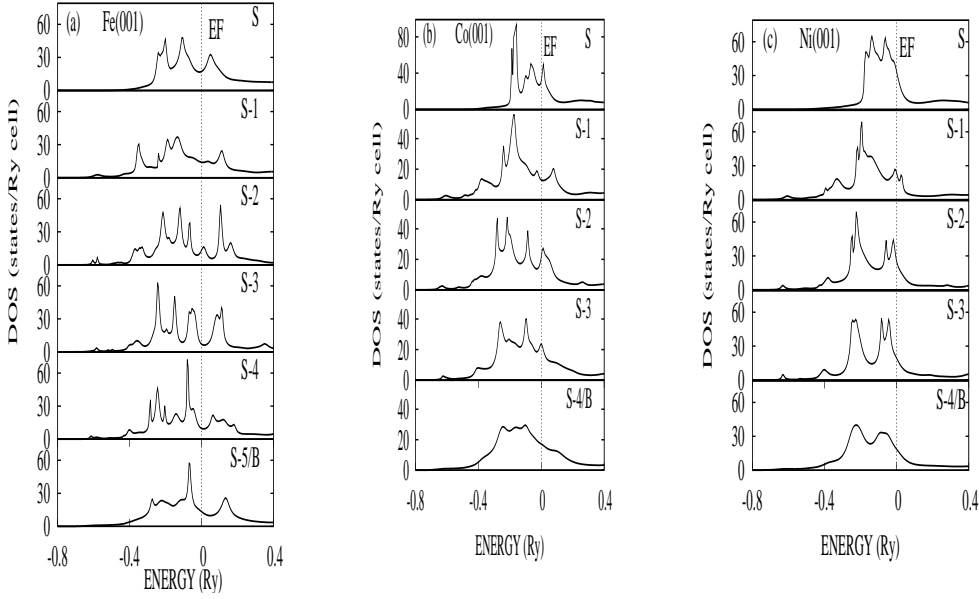}
\caption{Comparison of layer based local DOS for the surface (S), sub-surfaces (S-1, S-2, S-3, S-4, S-5) and bulk (B).(left) Fe (centre) Co (right) Ni.}
\end{figure}

The layer based total DOS, shown in the figure 3 indicates that  DOS approaches the bulk value from $5^{th}$ layer down the top most layer in the case of Fe(001) where as it approaches to the bulk from $4^{th}$ layer in the other two, Co(001) and Ni(001), cases. Therefore surface properties approach to the bulk properties after four layers down in the case of Fe(001) and after three layers in the cases of Co(001) and Ni(001). 

\begin{figure}[h]
\centering
\includegraphics[width=16cm,height=14cm]{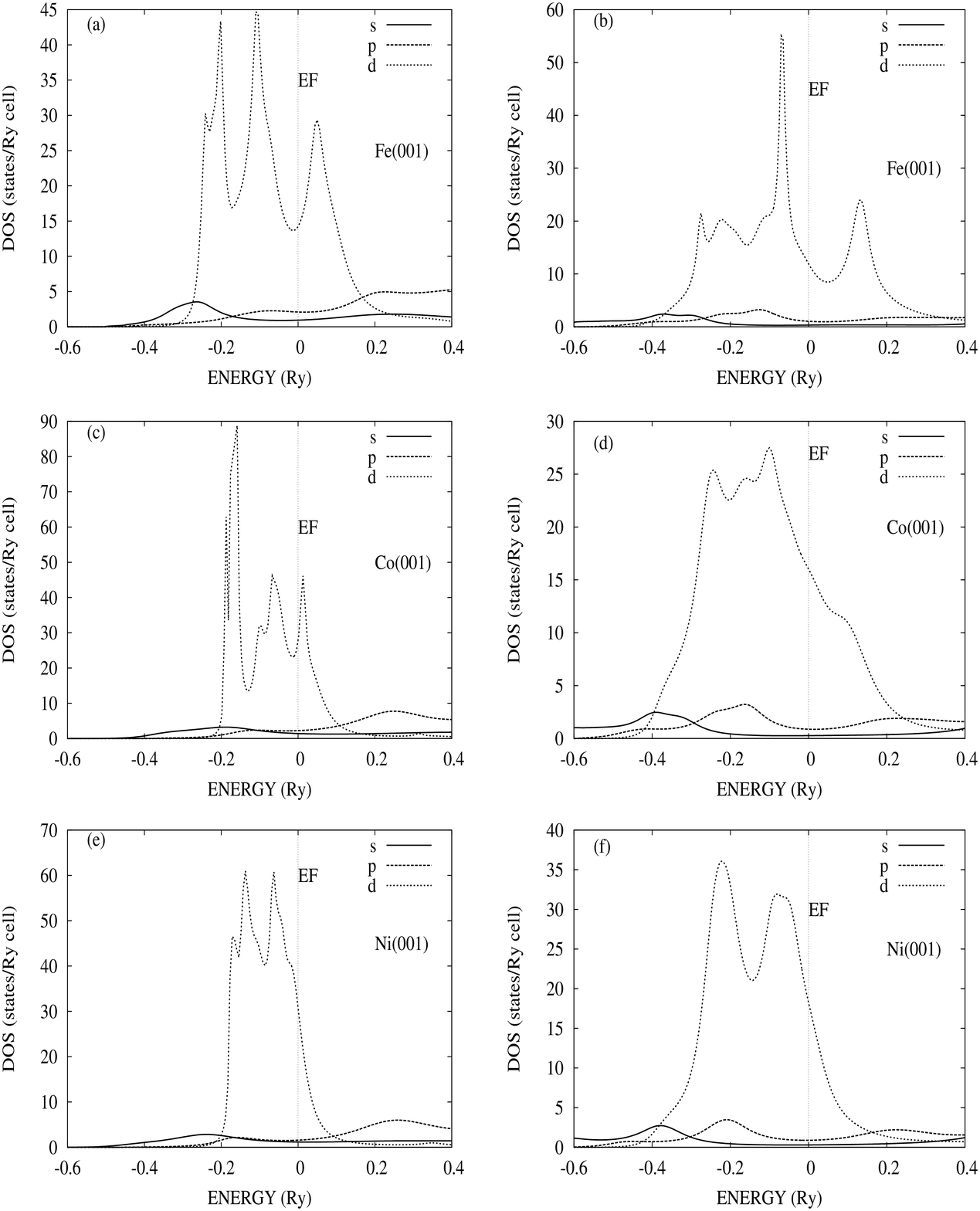}
\caption{Orbital resolved DOS: surface : (a), (c) \& (e) and bulk : (b), (d) \& (f)}
\end{figure} 

In transition metals, d-electrons contributes mainly to the formation of bands. Therefore, any surface effects must be reflected into d-band. This is evident from the figure 4, the orbital resolved DOS. The narrowing of d-band is due to the dehybridization of s-, p-, and d-electrons \cite{freeman2}. It is clear that for surface as well as for bulk, the DOS of s- and p-orbitals are negligible but that of d-orbitals is significant. Hence d-electrons give significant contribution towards the magnetic moment than that of s- and p-orbitals. This is expected. Because of narrowing of peaks, states are more localized and peak height is more. 

\begin{table}
\centering
{\small
\caption{Orbital resolved magnetic moment for surface (S), sub-surfaces (S-1, S-2, S-3, S-4) and bulk (B) in $\mu_B$/atom.}
\begin{tabular}{||c||c|c|c|c||c|c|c|c||c|c|c|c||}
\hline 
& \multicolumn{4}{|c||}{Fe(001)}  & \multicolumn{4}{|c||}{Co(001)} &\multicolumn{4}{|c||}{Ni(001)} \\ \cline{2-13}
\multicolumn{1}{||c||}{\raisebox{1.2ex}[0pt]{Layers}} & s & p & d & Total & s & p & d & Total & s & p & d & Total \\ \hline
S & 0.01 & 0.02 & 2.79 & 2.82 & -0.01 & 0.02 & 1.83 & 1.84& -0.01 & 0.0 & 0.70 & 0.69\\\hline 
S-1 & -0.02  & -0.03 & 2.13 & 2.08 & -0.01 & -0.03 & 1.49 & 1.45 & -0.01 & -0.01 & 0.60 & 0.58\\\hline 
S-2 & -0.03 & -0.06 & 2.06 & 1.97& -0.01 & -0.04 & 1.79 & 1.74& 0.0 & -0.02 & 0.62 & 0.60\\\hline 
S-3 & -0.02 & -0.04 & 2.70 & 2.64 & -0.01 & -0.06 & 1.66 & 1.59& 0.0 & -0.02 & 0.56 & 0.54 \\\hline 
S-4 & -0.01 & -0.06 & 2.43 & 2.36&&&&&&&&\\\cline{1-5} 
S-5/B & -0.02 & -0.06 & 2.25 & 2.17 & {\raisebox{1.2ex}[0pt]{-0.02}} & {\raisebox{1.2ex}[0pt]{-0.06}} & {\raisebox{1.2ex}[0pt]{1.62}} & {\raisebox{1.2ex}[0pt]{1.54}}& {\raisebox{1.2ex}[0pt]{-0.01}} & {\raisebox{1.2ex}[0pt]{-0.02}} & {\raisebox{1.2ex}[0pt]{0.55}} & {\raisebox{1.2ex}[0pt]{0.52}}\\\hline 
\end{tabular}
} 
\end{table}

The orbital resolved layer based magnetic moment of different layers and the bulk, tabulated in table 1, shows magnetic moment approaches to the bulk value from $5^{th}$ layer down  in case of Fe(001) and from  $4^{th}$ layer in the other two cases as expected from DOS result. Magnetic moments for different layers exhibit Friedel oscillation. This is also shown in the figure 5, layer based percentage change in magnetic moment compared to the bulk value. Figure 5 further shows that the enhancement of surface magnetic moment is same for Fe(001) and Ni(001) but it is less for Co(001). Table 1 also shows that d-electrons contribute most to the magnetic moment compared to s and p electrons as  expected form DOS. Figure 2 shows that at the Fermi level, difference between surface spin-up DOS and that of bulk is very small for Fe(001) and Co(001) and negligible for Ni(001). But there is significant difference between spin-down surface  DOS and that of bulk DOS. This shows spin-down states are mainly responsible for the enhancement of magnetic moment at the surface. In all the three cases, the contribution of s- and p-orbital is comparable to TB-LMTO Green's function method \cite{alden}. 

For comparison we have summarized our results of magnetic moments along with  other available calculated and experimental values in table 2. Table 2 shows that the enhancement of surface magnetic moment compared to the bulk is 30\% for Fe(001), 19\% for Co(001) and 32\% for Ni(001). Our result of bulk magnetic moment for Fe is within 5\% difference from experimental value \cite{danan, pauling}.  Since enhancement of surface magnetic moment is less for Co(001) than the other two, therefore it is more stable to the change in the environment.

\begin{table}
{\centering
{\small
\begin{tabular}{|c|c|c|c|c|c|c|c|c|c|c|c|c|c|c|}\hline 
& \multicolumn{4}{|c|}{Fe(001)}  & \multicolumn{5}{|c|}{Co(001)} &\multicolumn{5}{|c|}{Ni(001)} \\ \cline{2-15}
\multicolumn{1}{|c|}{\raisebox{1.2ex}[0pt]{Methods}} & S & S-1 & S-2 & C/B & S & S-1 & S-2 & S-3 & S-4/B & S & S-1 & S-2 & S-3 & S-4/B\\ \hline

 & 2.98 & 2.35& 2.39 & 2.20 &1.86  & 1.64 & 1.65 & 1.64 & 1.65 & 0.68 &&&0.56 & \\ 
       & \cite{ohnishi,freeman,wu}&\cite{ohnishi}&\cite{ohnishi}&\cite{ohnishi}&\cite{li}&\cite{li}&\cite{li} & \cite{li}&\cite{li}&\cite{freeman} &&&
\cite{freeman} & \\ 
\cline{2-15}
FP-  &2.80 & 2.38 & 2.43 &  2.15&&&&&&0.73&0.68& 0.66& 0.63&\\
LAPW  &\cite{freeman1} & \cite{wu} & \cite{wu} & \cite{freeman}&&&&&&\cite{wu} &\cite{wu} & \cite{wu}& \cite{wu}&\\
\cline{2-15}

  &&&& 2.30  &&&&&& 0.68 & 0.60 & 0.59 & 0.56 &\\
  &&&& \cite{wu} &&&&&& \cite{wimmer} & \cite{wimmer}& \cite{wimmer} &  \cite{wimmer}&\\
\hline
LMTO  & 2.87 & 2.34 & 2.33 & 2.18 &&&&&& 0.59 & 0.58 & 0.57 & 0.55 &\\
&\cite{eriksson,eriksson1}&\cite{eriksson,eriksson1}&\cite{eriksson,eriksson1}&\cite{eriksson,eriksson1}&&&&&&\cite{eriksson,eriksson1}&\cite{eriksson,eriksson1}&\cite{eriksson,eriksson1}&\cite{eriksson,eriksson1}& \\
\hline 
 & 2.97 & 2.30 & 2.37 & 2.24 & 1.84 & 1.63 & 1.66 & 1.64 & &0.69 & 0.64 & 0.66 & 0.64 & \\ 
 & \cite{alden} & \cite{alden} & \cite{alden} & \cite{alden} &  \cite{alden}& \cite{alden} &  \cite{alden}& \cite{alden} & &\cite{alden} &  \cite{alden}&\cite{alden} &  \cite{alden}& \\ 
\cline{2-15}
 & 2.97 & 2.30  & 2.37  & 2.25 & 1.84& 1.63&1.66&1.65&1.66&0.69& 0.64&0.66&0.64&0.65\\
 & \cite{niklasson}& \cite{niklasson} & \cite{niklasson} & \cite{niklasson} & \cite{niklasson}& \cite{niklasson}&\cite{niklasson}&\cite{niklasson}&\cite{niklasson}&\cite{niklasson}& \cite{niklasson}&\cite{niklasson}&\cite{niklasson}&\cite{niklasson}\\
\cline{2-15}
TB-  & 2.86& 2.16& 2.38& 2.17& 1.76 & 1.46 & 1.58 & 1.56& 1.58  & 0.65 & 0.53 & 0.61 & 0.60 & 0.59 \\
LMTO  & \cite{biplab}& \cite{biplab}& \cite{biplab}& \cite{biplab}& \cite{monodeep} & \cite{monodeep} & \cite{monodeep} & \cite{monodeep}& \cite{monodeep}  & \cite{monodeep}& \cite{monodeep} & \cite{monodeep} & \cite{monodeep} & \cite{monodeep}\\
\cline{2-15}
  & 2.98      & 2.17        &  2.40      & 2.26      &&&&&&&&&&\\ 
 &\cite{huda}& \cite{huda} & \cite{huda}&\cite{huda}&&&&&&&&&&  \\ 
\cline{2-15}
 & 2.95 &2.20 & 2.39  & 2.28  & &&& &&&&&&\\
 & \cite{monodeep} &\cite{monodeep} & \cite{monodeep}  & \cite{monodeep}  & &&& &&&&&&\\
\cline{2-15}
 & 2.99 &2.21 & 2.38  & 2.26  & &&& &&&&&&\\
 & \cite{huda} &\cite{huda} & \cite{huda}  & \cite{huda}  & &&& &&&&&&\\ 
 \hline 
LCAO & 3.01& 1.69& 2.13& 1.84&&&&&& 0.44 & 0.58 & 0.62 & 0.56&0.54 \\
 & \cite{wang}& \cite{wang}& \cite{wang}& \cite{wang}&&&&&& \cite{wang1} & \cite{wang1} & \cite{wang1} & \cite{wang1}&\cite{wang1} \\
\hline
{\bf ASR} & 2.99 & 2.17 & 2.38& 2.27 &&&&&&&&&& \\ 
 & \cite{biplab} & \cite{biplab} & \cite{biplab}& \cite{biplab} &&&&&&&&&& \\ 
\hline 
 & & & & 2.21&&&&& 1.71&&&&& 0.616 \\
{\bf EXPT} & & & & \cite{danan}&&&&&\cite{pauling} &&&&&\cite{danan}  \\
 \cline{2-15}
 &&&& 2.22&&&&&&&&&&\\
 &&&& \cite{pauling}&&&&&&&&&&\\
\hline
{\bf OUR} & 2.82 & 2.08 & 1.97 & 2.17& 1.84 & 1.45 & 1.74 & 1.59 & 1.54& 0.69 & 0.58 & 0.60 & 0.54 & 0.52\\ 
{\bf WORK} & &  &  & &  &  &  &  & &  &  &  &  & \\ 
\hline 
\end{tabular} 
\caption{Comparision of magnetic moment in $\mu_B$/atom for surface (S), sub-surfaces (S-1, S-2, S-3) and central layer or bulk (C/B). Number in the square brackets represents the reference numbers. Note that,
Refs \cite{wimmer,alden,niklasson} use Green's function techniques;
Ref \cite{biplab} is with TB-LMTO including surface dilatation;
Refs \cite{huda,biplab} are Supercell calculation;
Refs \cite{huda,monodeep} use Real space recursion.}
}}
\end{table}

\begin{figure}[h]
\centering
\includegraphics[width=7cm,height=9cm,angle=-90]{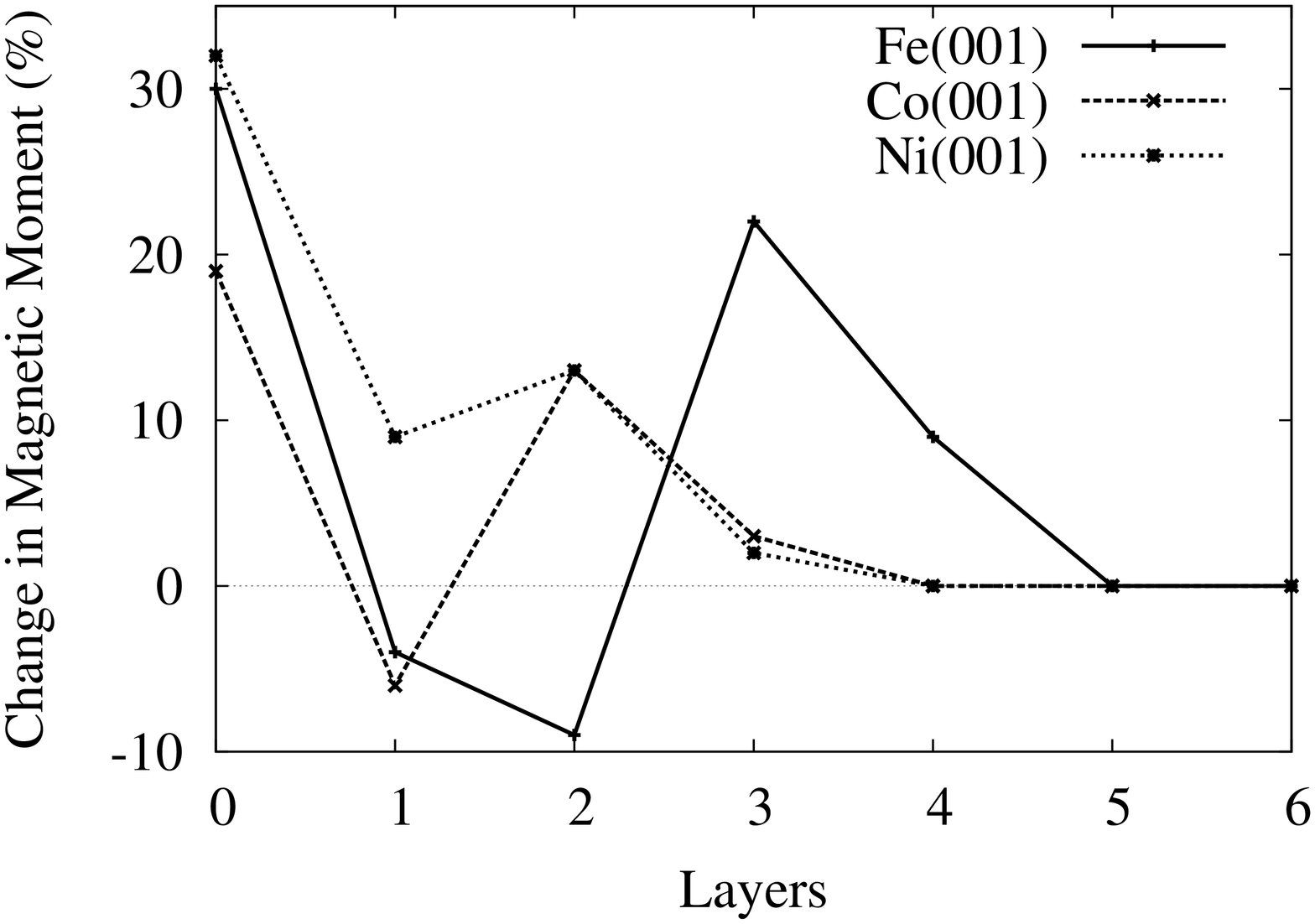}
\caption{Percentage variation of magnetic moment with respect to the bulk value for different layers}
\end{figure}
Experimental values of local magnetic moment, like surface layer and layers below are not available for comparison. It is only the bulk magnetic moment for which experimental result is available. Work function is a surface property and is related to the surface DOS. It is an experimentally measurable quantity and therefore theoretical calculation of it is necessary for comparison with experimental result to test the  accuracy of theoretical study. Table 3 shows our calculated work functions for all the systems under consideration agree quite well with experimental and other theoretical calculations.\\

\begin{table}
\caption{Work functions of Fe(001), Co(001) and Ni(001) in eV. Number in the square brackets represents the reference numbers.}
\begin{center}
\begin{tabular}{|c|c|c|c|}
\hline 
Methods & Fe (001) & Co(001) & Ni(001) \\ \hline
FPLAPW &4.29 \cite{ohnishi} & 5.17 \cite{li} & 5.37 \cite{wimmer}, 5.5 \cite{freeman2},5.31 \cite{ohwaki}  \\ \hline
LAPW & & & 5.71 \cite{ingle}\\ \hline
LMTO &4.30 \cite{eriksson1} &  & 5.02 \cite{eriksson1}\\ \hline
TB-LMTO & 4.5 \cite{alden} & 5.52 \cite{alden} & 5.75 \cite{alden}\\ \hline
Experimental &4.67 \cite{michael}, 4.4 \cite{tuner} &5.0 \cite{michael}  & 5.22 \cite{michael}\\ \hline
Our work &4.08 &5.30  &4.76 \\ \hline
\end{tabular} 
\end{center}
\cite{alden}: with Green's function.\\
\cite{freeman2}, \cite{ohwaki}, \cite{ingle}: with surface embedded Green's function.
\end{table}

\section{Conclusion}
Augmented space formalism coupled with recursion method \& TB-LMTO is applied for the calculation of layer dependent density of states and magnetic moment for bcc Fe(001), fcc Co(001) and fcc Ni(001). It is found that the magnetic moment get enhanced at the surface compared to the bulk value by 30\%, 19\% and 32\% for Fe(001), Co(001) and Ni(001) respectively. Layer wise magnetic moments show Friedel oscillations. This observation and calculated magnetic moments agree well with the available experimental result and other calculations. The enhancement of surface magnetic moment is due to narrowing down of d-band, which support earlier studies. It is observed that bulk magnetic moment is attained at the $5^{th}$ layer down the top most surface layer in the case of Fe(001) and at the  $4^{th}$ layer in the cases of Co(001) and Ni(001). The calculated work function is 4.08 eV, 5.30 eV and 4.76 eV for Fe(001), Co(001) and  Ni(001) respectively. These values are closed to the available experimental and other calculated results.\\
Since Augmented Space Recursion method is real space technique, therefore it is not only suitable for the study of local properties like surface and different layers, it can also be applied to cases where surface is rough and interface layers between any substrate and a film which are normally disordered due to inter diffusion. 

\section*{Acknowledgments}
Authors would like to acknowledge Defense Research and Development Organization, India for providing financial support to this work under grant number ERIP/ER/1006016/ M/01/1373. P Parida acknowledges INSPIRE program division, Department of Science and Technology, India for providing a Senior Research Fellowship. We would like to thank Prof. O.K. Anderson, Max Plank Institute, Stuttgart, Germany, for his kind permission to use TB-LMTO code developed by his group.

\bibliographystyle{unsrt}
\bibliography{final}

\end{document}